\documentclass[twoside]{article}
\usepackage{amsmath}
\usepackage{amsfonts}
\usepackage{amssymb}

\usepackage{qic,epsfig}

\usepackage{bbm}
\usepackage{color}

\usepackage{enumerate}
\usepackage{mathtools}
\mathtoolsset{showonlyrefs}

\usepackage{tikz}
\usetikzlibrary{backgrounds,calc,trees,hobby,decorations.pathreplacing}

\newtheorem{remark}{Remark}
\newtheorem{example}{Example}

\newcommand{\tr}{\operatorname{tr}}
\newcommand{\id}{\mathbbm{1}}

\newcommand{\kernel}{\operatorname{ker}}
\newcommand{\diag}{\operatorname{diag}}
\newcommand{\rank}{\operatorname{rank}}

\newcommand{\R}{\mathbb{R}}
\newcommand{\C}{\mathbb{C}}
\newcommand{\im}{\operatorname{im}}
\newcommand*{\rhop}{\rho^{\prime}}

\newcommand*{\Up}{{U^{\prime}}}
\newcommand*{\cH}{\mathcal{H}}
\newcommand*{\cB}{\mathcal{B}}

\begin{document}
\setlength{\textheight}{8.0truein}    

\runninghead{On quantum additive Gaussian noise channels}
            {Martin Idel and Robert K\"onig}

\normalsize\textlineskip
\thispagestyle{empty}
\setcounter{page}{1}

\copyrightheading{17}{3\&4}{2017}{0283--0302}

\vspace*{0.88truein}

\alphfootnote

\fpage{1}

\centerline{\bf
On quantum additive Gaussian noise channels}
\vspace*{0.37truein}
\centerline{\footnotesize
Martin Idel}
\vspace*{0.015truein}
\centerline{\footnotesize\it Zentrum Mathematik, Technische Universit\"at M\"unchen,}
\baselineskip=10pt
\centerline{\footnotesize\it  85748 Garching, Germany}
\vspace*{10pt}
\centerline{\footnotesize 
Robert K\"onig}
\vspace*{0.015truein}
\centerline{\footnotesize\it Institute for Advanced Studies \& Zentrum Mathematik}
\baselineskip=10pt
\centerline{\footnotesize\it  Technische Universit\"at M\"unchen, 85748 Garching, Germany}
\vspace*{0.225truein}
\publisher{August 22, 2016}{February 13, 2017}

\vspace*{0.21truein}

\abstracts{
We give necessary and sufficient conditions for a  Gaussian quantum channel to have a dilation involving a passive, i.e., number-preserving unitary. We then establish a normal form of such channels: any passively dilatable channel is the result of applying passive unitaries to the input and output of a Gaussian additive channel. The latter combine the state of the system with that of the environment by means of a  multi-mode beamsplitter.  }{}{}

\vspace*{10pt}

\vspace*{3pt}

\vspace*{1pt}\textlineskip    

\section{Introduction} 
It is a fortunate fact of nature that many physical systems are well-described by a quadratic approximation. Harmonic oscillators are ubiquitous in physics, and are the basis for our understanding of a variety of phenomena in the domain of classical mechanics, electrodynamics, solid state physics, quantum field theory and gravity. Gaussian processes are also essential in probability theory and information theory as a source of non-trivial yet exactly solvable scenarios of interest. Arguably one of the most promiment examples is Shannon's capacity formula for the additive white Gaussian noise (AWGN) channel~\cite{Shannon48}. The latter constitutes a realistic model for fiberoptic communication. It transforms an analog input  signal~$X$ (modeled by a random variable on~$\mathbb{R}^n$) into the output $Y=X+Z$ by adding  an independent centered unit-variance Gaussian random variable~$Z$ representing the noise.  More generally,~$Z$ may be replaced by an arbitrary random variable~$Z$, in which case we refer to this as an additive noise channel.

In quantum mechanics, Gaussian states arise naturally as thermal states of  Hamiltonians which are quadratic in the mode operators of a bosonic system. The latter  provide an accurate description of many systems of interest. Restricting to such  Hamiltonians, Gaussian channels result whenever a system interacts with an environment in a Gaussian state. A typical example is a  channel of the form 
\begin{align}
\mathcal{E}(\rho)=\tr_E\left(U_\lambda(\rho\otimes\rho_E)U_\lambda^*\right)\ ,\label{eq:additivenoisece}
\end{align} where $U_\lambda$ is a beamsplitter with transmissivity~$\lambda\in [0,1]$, and $\rho_E$~is a Gaussian state of the environment (see Example~\ref{ex:beamsplitterexample} below). This channel constitutes a natural quantum counterpart of the classical additive noise channel, and, correspondingly, we refer to it as a (quantum) {\em additive Gaussian noise channel}. In the special case where~$\rho_E$ is the thermal state of the harmonic oscillator Hamiltonian, it is also called a thermal noise channel (and is the counterpart of the AWGN channel).

The channel~\eqref{eq:additivenoisece} also arises naturally 
from the viewpoint of resources in e.g., quantum optics. The  unitary~$U_\lambda$ obeys a special property: it cannot generate squeezing. More generally, a unitary~$U$ acting jointly on $n$~modes of a system and $l$~environment modes is called {\em passive} if it commutes with the total number operator~$\hat{N}=\sum_{k=1}^{n+l} a_k^* a_k$.  Here $a_k=(Q_k+iP_k)/\sqrt{2}$ is the usual annihilation operator associated with the~$k$-th mode.  The unitary~$U_\lambda$ describing the beamsplitter is an example of such a passive unitary. In fact, a Gaussian unitary is passive if and only if it is  the composition of beamsplitters and phase shifters~\cite{reckzeilinger}. Thus passive Gaussian unitary operations are experimentally easy to implement. Physically, such operations neither introduce nor remove photons and are thus implementable without expending energy.

Considering squeezing as a resource, it is natural to try to separate preexisting squeezing (in the form of a potentially squeezed state of the environment) from evolutions generating squeezing. One is then led to consider the class of {\em passively dilatable channels}: these are channels possessing a dilation with a passive unitary. Motivated by the decomposition~\cite{reckzeilinger} of passive Gaussian unitaries, we  ask if passively dilatable channels also have a special structure. The main result of our paper is such a normal form: we establish a close connection between additive channels and the class of passively dilatable channels. That is, any passively dilatable channel is the composition of 
 (i)~a passive unitary applied to the input, (ii) an additive Gaussian noise channel and (iii)~a passive unitary applied to the output. 
 
 Our result thus provides an alternative characterization of quantum additive channels as canonical examples of non-unitary channels which do not generate squeezing.  It is a further manifestation, but in a non-unitary context, of the well-known fact that non-linear optical elements are generally required for the generation of squeezed states~\cite{braun05}. We refer to~\cite{idellercherwolf} for a recent study of the operational quantification of squeezing, and a more detailed discussion of its role in quantum optics.
 
 Our work also establishes simple necessary and sufficient criteria for deciding when a given passively dilatable channel has a dilation with~$l$ environment modes. Our considerations cover all cases, including rank-deficient ones. Using these criteria, we compute the minimal number of required environment modes for a passive dilation to exist. These results are similar, in spirit, to those of~\cite{car08,careigioho11}, but  in contrast to the latter, geared towards characterizing non-squeezing resources. Specifically,~\cite{car08} constructs a unitary dilation of an arbitrary Gaussian quantum channel, and presents a number of applications to weak degradability.  In~\cite{careigioho11}, the minimal number of environment modes required to provide a unitary Gaussian dilation with pure state environment is identified, and bounds for the case of mixed state enviroments are given (see Remarks~\ref{rem:mixeddilation} and~\ref{rem:puredilations} below).

\section{Preliminaries}
We begin by introducing some of the basic relevant terminology associated with continuous variable quantum information (for longer reviews of the material see for instance~\cite{eis07,weedbrock2012}). This will also serve to introduce our notation.
\subsection{Gaussian states and operations\label{sec:gaussianoperations}}
We consider $n$-mode bosonic systems with $n$ pairs of quadratures  (or modes) given by~$R=(Q_1, P_1, Q_2, P_2,\allowbreak \ldots, Q_{n}, P_{n})$, or, equivalently,
the annihilation and creation operators
\begin{align}
a_k=\frac{1}{\sqrt{2}}(Q_k+i P_k)\qquad\textrm{ and }\qquad a_k^*=\frac{1}{\sqrt{2}}(Q_k-iP_k)
\end{align}
for $k=1,\ldots,n$. 
The commutators 
\begin{align} 
[R_j,R_k]=i\sigma_{jk}\mathsf{id}\label{eq:ccrrelations}
\end{align} are given by the standard symplectic form
\begin{align}
	\sigma:=\bigoplus_{i=1}^n \begin{pmatrix}{} 0 & 1 \\ -1 & 0 \end{pmatrix}. \label{eqn:sympform}
\end{align}
To simplify notation, it is often convenient to work in the permuted basis~$(Q_1,\ldots,Q_m,P_1,\allowbreak\ldots,P_m,Q_{m+1},\ldots,Q_{m+l},P_{m+1},\ldots,P_{m+l})$, where $\sigma$ takes the form $\sigma=\sigma_{2m}\oplus \sigma_{2l}$ with
\begin{align*}
	\sigma_{2k}:=\begin{pmatrix}{} 0_{k\times k} & \id_k \\ -\id_k & 0_{k\times k} \end{pmatrix}.
\end{align*}
For concreteness, we will henceforth  assume that the CCR-relations~\eqref{eq:ccrrelations} are realized by unbounded operators
acting on the tensor product~$\cH^{\otimes n}$ where $\cH\cong L^2(\mathbb{R})$ is the Hilbert space associated with a single mode. When convenient, we will also use the notation $\cH_{A_1\cdots A_n}=\cH^{\otimes n}$ to denote multipartite Hilbert spaces. 

An important subset of states is given by the {\em Gaussian states}: such a state~$\rho$ is fully  characterised by its first and second moments
\begin{align}
d_k &=\tr(\rho R_k)\qquad\textrm{ and }\qquad \gamma_{k\ell}=\tr(\rho\{R_k-d_k\mathsf{id},R_\ell-d_{\ell }\mathsf{id}\})\ ,
\end{align}
where $\{A,B\}=AB+BA$ denotes the anticommutator.    Here $d\in\mathbb{R}^{2n}$ is the {\em displacement vector}, whereas the symmetric matrix  $\gamma=\gamma^T\in\mathbb{R}^{2n\times 2n}$ is referred to as the {\em covariance matrix}. By Heisenberg's uncertainty principle, the covariance matrix of any state satisfies the operator inequality
\begin{align}
\gamma\geq i\sigma_{2n}\label{eq:operatorinequalitycovariance}\ .
\end{align}
Conversely, any pair $(d,\gamma)$ with $d\in\mathbb{R}^{2n}$ and $\gamma=\gamma^T\in\mathbb{R}^{2n\times 2n}$ satisfying~\eqref{eq:operatorinequalitycovariance} uniquely defines a Gaussian $n$-mode state.\footnote{We emphasize, however, that passive dilatability (as defined below) of the channel $(X, Y, 0^{2n})$
is not equivalent to passive dilatability of the channel $(X, Y, v)$ with $v \neq 0^{2n}$. For example, in
the case of pure translations $(X, Y) = (\id_{2n}, 0_{2n\times 2n})$, the corresponding unitary channel is passively
dilatable if and only if $v = 0^{2n}$.} As a consequence, we may identify the set of Gaussian states with the set of such pairs.
 
\subsection{Gaussian operations}
A quantum operation (or channel)  acting on an $n$-mode system is described by a  completely positive trace-preserving map~$\Phi:\cB(\cH^{\otimes n})\rightarrow\cB(\cH^{\otimes n})$. Here $\cB(\cH^{\otimes n})$ is the set of bounded linear operators on~$\cH^{\otimes n}$. Again, the subset of {\em Gaussian channels} is distinguished by the property that such channels map Gaussian states to Gaussian states. 
Such a channel is completely characterized by its action on Gaussian states, and the latter has a convenient description: for a Gaussian state~$\rho$ with displacement vector~$d$ and covariance matrix~$\gamma$, the Gaussian state~$\Phi(\rho)$ resulting from application of the channel is described by the pair $(d',\gamma')$ obtained from the map
\begin{align*}
	\gamma &\mapsto X\gamma X^T+Y \\
	d&\mapsto Xd+v\ ,
\end{align*}
where  the matrices~$X,Y\in\mathbb{R}^{2n\times 2n}$ and the vector $v\in\mathbb{R}^{2n}$ determine the action of the channel. Clearly, $Y=Y^T$ has to be symmetric for this to map covariance matrices to covariance matrices. The map is completely positive if and only if\footnote{Note that in \cite{eis07}, the condition is stated with a minus sign, but since $\sigma^T=-\sigma$ and $Y$ is symmetric, this conditions is equivalent.} (cf. \cite{eis07})
\begin{align}
	Y\geq i\sigma_{2n}-iX\sigma_{2n} X^T\ . \label{eqn:comppos}
\end{align}
Conversely, and similarly as for Gaussian states, any triple $(X,Y,v)$ with~$Y=Y^T$ symmetric, $(X,Y)$ satisfying~\eqref{eqn:comppos} and $v\in\mathbb{R}^{2n}$ arbitrary uniquely determines a Gaussian $n$-mode channel. We will thus identify the set of Gaussian channels with
the set of such triples. 

In fact, the displacement vector~$v\in\mathbb{R}^{2n}$  has no 
influence on operational properties of the channel such as capacities since it can be changed arbitrarily by applying a displacement operator (a Gaussian, but non-passive unitary) to the output of the channel (see e.g.,~\cite{eis07}). In contrast, the matrices~$(X,Y)$ determine all important characteristics of the channel. As a consequence, we will henceforth assume that $v=0$ (as in~\cite{car08,careigioho11}), and write
$\Phi_{X,Y}:\cB(\cH^{\otimes n})\rightarrow\cB(\cH^{\otimes n})$ for the Gaussian channel determined by the pair~$(X,Y)$.

\subsection{Gaussian unitaries and passive unitaries}
A Gaussian unitary channel is one of the form $\Phi_{X,0}$ (i.e., $Y=0$). 
For such channels, the constraint~\eqref{eqn:comppos} implies that~$X$ preserves the symplectic form (i.e., $X\sigma_{2n} X^T=\sigma_{2n}$), i.e., $X$ is an element of
\begin{align}
Sp(2n)=\{S\in \R^{2n\times 2n}\ |\ S\sigma_{2n} S^T=\sigma_{2n}\}\ ,
\end{align}
the group of real symplectic matrices. It can be shown that any element~$S\in Sp(2n)$ defines a unitary $U_S$ on $\cH^{\otimes n}$ such that 
\begin{align}
\Phi_{S,0}(\rho)=U_S\rho U^*_S\ .
\end{align}
Furthermore, $S\mapsto U_S$ defines a representation called the metaplectic representation of~$Sp(2n)$. 

In more physical terms, a Gaussian unitary describes the evolution (for a fixed amount of time) generated by a Hamiltonian~$H$ which is quadratic in the creation- and annihilation operations, i.e., one that has the form
\begin{align}
H=\sum_{j,k=1}^n h_{j,k}a_j^* a_k+h.c.\label{eq:quadratichamiltonians}
\end{align}

A Hamiltonian of the form~\eqref{eq:quadratichamiltonians} which  commutes with the total number operator, i.e., satisfies
\begin{align}
[H,\sum_{j=1}^n a_j^* a_j]=0
\end{align}
is called \emph{passive}. A passive Hamiltonian generates Gaussian unitaries which are associated with orthogonal symplectic matrices $S\in Sp(2n)\cap O(2n)$, where
\begin{align}
O(2n)&=\{O\in \mathbb{R}^{2n\times 2n}\ |\ OO^T=\id_{2n}\}\ .
\end{align}
We call such Gaussian unitaries {\em passive}. 
It can be shown that passive Gaussian unitaries can be realized using beamsplitters and phase shifters only~\cite{reckzeilinger}. 
   
\subsection{On the orthogonal symplectic group}
Let us collect a few facts about the group $Sp(2n)\cap O(2n)$. 
Crucially, there is an isomorphism $U(n)\cong Sp(2n)\cap O(2n)$  between
 this group and the 
group
\begin{align}
U(n)&=\{U\in\mathbb{C}^{n\times n}\ |\ U^\dagger U=\id_{n}\}\ 
\end{align}
of unitary $n\times n$~matrices. For our purposes, it will be convenient to write out this isomorphism for the case of $n+l$~modes (associated with a system and its environment), as follows: 

\begin{lemma} \label{lem:unitisom}
The map
\begin{align}
\begin{matrix} \phi: U(n+l) & \rightarrow & Sp(2(n+l))\cap O(2(n+l))\\
 U=\begin{pmatrix}{} u_1 & u_2 \\ u_3 & u_4 \end{pmatrix} & \mapsto & 
\phi(U)=S=\begin{pmatrix}{} s_1 & s_2 \\ s_3 & s_4 \end{pmatrix} 
\end{matrix}\qquad\textrm{ where }\quad 
s_i=\begin{pmatrix}{} \operatorname{Re}(u_i) & \operatorname{Im}(u_i) \\ -\operatorname{Im}(u_i) & \operatorname{Re}(u_i) \end{pmatrix} \label{eqn:isom}
\end{align}
and where $u_1\in \mathbb{C}^{n\times n}, u_2\in \mathbb{C}^{n\times l},
u_3\in \mathbb{C}^{l\times n}, u_4\in \mathbb{C}^{l\times l}$ is an isomorphism.
\end{lemma}
\proof{The existence of the isomorphism is well-known (see \cite{mcd98}). We need to show that $S$ is symplectic:
\begin{align*}
	S\sigma_{2(n+l)}S^T=\begin{pmatrix}{} s_1\sigma_{2n}s_1^T+s_2\sigma_{2l}s_2^T & s_1\sigma_{2n}s_3^T+s_2\sigma_{2l}s_4^T \\ s_3\sigma_{2n}s_1^T+s_4\sigma_{2l}s_2^T & s_3\sigma_{2n}s_3^T+s_4\sigma_{2l}s_4^T \end{pmatrix}
\end{align*}
Note that
\begin{align*}
	s_i\sigma_{2n}s_j^T=\begin{pmatrix}{} \operatorname{Re}(u_i)\operatorname{Im}(u_j)^T-\operatorname{Im}(u_i)\operatorname{Re}(u_j)^T & \operatorname{Re}(u_i)\operatorname{Re}(u_j)^T+\operatorname{Im}(u_i)\operatorname{Re}(u_j)^T \\ -(\operatorname{Re}(u_i)\operatorname{Re}(u_j)^T+\operatorname{Im}(u_i)\operatorname{Re}(u_j)^T) & \operatorname{Re}(u_i)\operatorname{Im}(u_j)^T-\operatorname{Im}(u_i)\operatorname{Re}(u_j)^T \end{pmatrix}
\end{align*}
and that 
\begin{align*}
	\operatorname{Re}(u_i)\operatorname{Re}(u_j)^T+\operatorname{Im}(u_i)\operatorname{Re}(u_j)^T=\operatorname{Re}(u_iu_j^{\dagger}) \\
	\operatorname{Re}(u_i)\operatorname{Im}(u_j)^T-\operatorname{Im}(u_i)\operatorname{Re}(u_j)^T=\operatorname{Im}(u_iu_j^{\dagger}).
\end{align*}
Therefore, since $U$ is unitary, it follows that $S\sigma_{2(n+l)}S^T=\sigma_{2(n+l)}$. Similarly, 
\begin{align*}
	SS^T=\begin{pmatrix}{} s_1s_1^T+s_2s_2^T & s_1s_3^T+s_2s_4^T \\ s_3s_1^T+s_4s_2^T & s_3s_3^T+s_4s_4^T \end{pmatrix}=\id_{2n}
\end{align*}
using the unitarity of $U$. To prove that this is an isomorphism, one then has to consider the inverse map. This is well-defined because any matrix $S\in Sp(2n)\cap O(2n)$ is of the form of the image of the map $\phi$ in (\ref{eqn:isom}) (see \cite{mcd98}).}

The following lemma will be an important tool in what follows. 
\begin{lemma} \label{lem:commute}
For any matrix $X\in \R^{2n\times 2n}$, $[X,\sigma_{2n}]=0$ if and only if $X$ has the form
\begin{align}
	X=\begin{pmatrix}{} A & B \\ -B & A \end{pmatrix}\label{eqn:eqcommutatorformsigma}
\end{align}
for some matrices $A,B\in \R^{n\times n}$. In particular, any matrix $X\in Sp(2n)\cap O(2n)$ commutes with~$\sigma_{2n}$. 

Furthermore, any eigenvalue of a matrix of form (\ref{eqn:eqcommutatorformsigma}) has even multiplicity.
\end{lemma} 
In fact, it can be shown (see~\cite{mcd98}) that any two of the three properties $X\sigma_{2n}X^T=\sigma$, $[X,\sigma_{2n}]=0$ and $XX^T=\id_{2n}$ implies the third, a feature known as the $2$-out-of-$3$~property. 

\proof{The proof is straightforward. The fact that this holds for $X\in Sp(2n)\cap O(2n)$ is clear from Lemma \ref{lem:unitisom} (specialized to $l=0$).

For the eigenvalue multiplicity, note that if $v\equiv (v_1,v_2)^T$ with $v_1,v_2\in \R^n$ is an eigenvector to the eigenvalue $\lambda$ of $X$, then $\sigma_{2n}v=(v_2,-v_1)^T$ is an eigenvector to the same eigenvalue and $\sigma_{2n}v\perp v$. Now, if $\{v,\sigma_{2n}v\}^{\perp}$ contains another eigenvalue $w\in \R^{2n}$ with eigenvalue $\lambda$, then $\sigma_{2n} w$ is again an eigenvector of $X$ with eigenvalue $\lambda$. We claim that $\{v,\sigma_{2n}v,w,\sigma_{2n}w\}$ is an orthonormal set of eigenvectors to eigenvalue $\lambda$. By construction, we have $w\perp \{v,\sigma_{2n} v\}$ and $\sigma_{2n}w\perp w$. Finally, $\sigma_{2n}w \perp v$ as $\langle \sigma_{2n}w,v\rangle=-\langle w,\sigma_{2n}v\rangle=0$. Iteratively, we can construct an orthonormal basis of every eigenspace, which will necessarily have even multiplicity.}

The next lemma is an extension theorem for orthogonal symplectic matrices:
\begin{lemma} \label{lem:extension}
Assume that $s_1\in \R^{2n\times 2n}$ and $s_2\in \R^{2n\times 2l}$ satisfy
\begin{align}
	\begin{split}
		s_1\sigma_{2n}s_1^T+s_2\sigma_{2l}s_2^T=\sigma_{2n} \\
		s_1s_1^T+s_2s_2^T=\id_{2n}\  . \label{eqn:orthogextend}
	\end{split}
\end{align}
Then there are  $s_3\in \R^{2l\times 2n}$ and $s_4\in \R^{2l\times 2l}$ such that 
\begin{align}
	S=\begin{pmatrix}{} s_1 & s_2 \\ s_3 & s_4 \end{pmatrix} \in Sp(2(n+l))\cap O(2(n+l))\ .\label{eq:ssymplecticall}
\end{align}
Furthermore, 
if $S$ is of the form~\eqref{eq:ssymplecticall} and
\begin{align}
	S^{\prime}=\begin{pmatrix}{} s_1 & s_2 \\ s^{\prime}_3 & s^{\prime}_4 \end{pmatrix} \in Sp(2(n+l))\cap O(2(n+l))\ ,
\end{align}
then there is an orthogonal symplectic matrix $o\in Sp(2l)\cap O(2l)$ such that 
\begin{align}
S^{\prime}&=
\begin{pmatrix}{}
\id_{2n} & 0_{2n\times 2l}\\
 0_{2l\times 2n} & o
 \end{pmatrix}S\ .\label{eq:sprimedef}
\end{align}
\end{lemma}
\proof{This is essentially saying that one can always extend suitable matrices to orthogonal symplectic matrices. It is clear by symplectic Gram-Schmidt (see \cite{mcd98}) that it is always possible to find $s_3,s_4$ to construct a symplectic matrix $S$, which however is not necessarily orthogonal. Therefore, we take the isomorphism to unitary matrices: Since $s_1$ and $s_2$ satisfy the relations~\eqref{eqn:orthogextend}, we can choose $u_1,u_2$ from the isomorphism in Lemma~\ref{lem:unitisom}. In particular, the matrix $V:=\begin{pmatrix}{} u_1 & u_2\end{pmatrix}$ fulfills $VV^{\dagger}=\id_{2n}$, hence we can extend it to a unitary matrix~$U$ and use the isomorphism again to find~$s_3$ and~$s_4$. The corresponding~$S$ is now orthogonal symplectic by construction.

For the second statement, let $S,S^{\prime}\in Sp(2(n+l))\cap O(2(n+l))$ be given by~\eqref{eq:ssymplecticall}
and~\eqref{eq:sprimedef}, respectively. Then 
\begin{align}
S^{\prime}S^T&=
\begin{pmatrix}{}
\id_{2n} & s_1s_3^T+s_2s_4^T\\
s_3^{\prime}s_1^T+s_4^{\prime}s_2^T & s_3^\prime s_3^T+s_4^\prime s_4^T
\end{pmatrix}\label{eq:SSprimeevaluated}
\end{align}
by the orthogonality  relation~\eqref{eqn:orthogextend}.
But $S^{\prime}S^T\in Sp(2(n+l))\cap O(2(n+l))$, hence it follows that
\begin{align}
S^{\prime} S^T&=
\begin{pmatrix}{}
 \id_{2n} & 0_{2n\times 2l}\\
 0_{2l\times 2n} & o
\end{pmatrix}=:O
\end{align} 
for some $o\in Sp(2l)\cap O(2l)$.  Combining this with~\eqref{eq:SSprimeevaluated} immediately gives 
$O^T S^{\prime }S^T=\id_{2(n+l)}$. The claim follows by left- and right-multiplying the latter identity with~$O$ and~$S$, respectively.}

\subsection{Dilations of Gaussian channels}
Consider the Gaussian $n$-mode channel~$\Phi_{X,Y}$ as defined in Section~\ref{sec:gaussianoperations}. It is well-known (see~\cite{car08}) that one can find a
 Gaussian state~$\rho_E$ of $l\leq n$~environment modes
and a Gaussian unitary matrix~$U$ acting on $n+l$~modes such that $\Phi_{X,Y}$  can be written as
\begin{align}
	\Phi(\rho)=\tr_E(U(\rho\otimes \rho_E)U^*)\ . \label{eqn:dilation}
\end{align}
Note that we do not demand~$\rho_E$ to be a pure state (if it is, this is referred to as the Stinespring representation,  see Remark~\ref{rem:puredilations} below). 
In Eq.~\eqref{eqn:dilation},  $U=U_S$ is the image under the metaplectic representation of a symplectic matrix $S\in Sp(2(n+l))$. 

The relationship between~$S$  and~$(X,Y)$ is obtained by analyzing the action on covariance matrices: if the $l$-mode Gaussian state $\rho_E$ has covariance matrix~$\gamma_E$, then the channel's action is given by
\begin{align*}
	\gamma\mapsto (S(\gamma\oplus \gamma_E)S^T)_{2n\times 2n}= X\gamma X^T+Y
\end{align*}
where $(\cdot)_{2n\times 2n}$ means that we restrict to the upper left block of the size $2n\times 2n$. More precisely, writing
\begin{align*}
	S=\begin{pmatrix}{} s_1 & s_2 \\ s_3 & s_4 \end{pmatrix}
\end{align*}
with $s_1\in \mathbb{R}^{2n\times 2n}$ and $s_4\in \mathbb{R}^{2l\times 2l}$, we have 
\begin{align}
	\begin{pmatrix}{} s_1 & s_2 \\ s_3 & s_4 \end{pmatrix}\begin{pmatrix}{} \gamma & 0_{2n\times 2l}\\0_{2l\times 2n} & \gamma_E\end{pmatrix}\begin{pmatrix}{} s_1 & s_2 \\ s_3 & s_4 \end{pmatrix}^T
	&=\begin{pmatrix}{} s_1\gamma s_1^T+s_2\gamma_Es_2^T & * \\ * & * \end{pmatrix} \label{eqn:extend}
\end{align}
and therefore 
\begin{align}
X\gamma X^T+Y=s_1\gamma s_1^T+s_2\gamma_Es_2^T\ \label{eq:actioncovariancematrix}
\end{align}
for all covariance matrices~$\gamma$. Thus the pair $(X,Y)$ and $(s_1,s_2,\gamma_E)$ are related by 
\begin{align}
X=\pm s_1\qquad\textrm{ and }\qquad Y=s_2\gamma_Es_2^T\ .\label{eq:covariancematrixxydef}
\end{align}

\section{Passively dilatable Gaussian channels}
Given a Gaussian channel $\Phi_{X,Y}$, we ask if there is a passive unitary associated with an element $S\in Sp(2n)\cap O(2n)$ and an (arbitrary) Gaussian state~$\rho_E$  of the environment constituting a dilation of the channel. We shall call any channel with this property~{\em passively dilatable}. Our main result is the following.
\begin{theorem} \label{thm:dilation}
Let $\Phi_{X,Y}$ be an $n$-mode Gaussian channel. The following conditions are equivalent:
\begin{enumerate}[(i)]
	\item There exists a passive dilation with $l$ environment modes and $S\in Sp(2(n+l))\cap O(2(n+l))$.\label{it:enumfirst}
	\item The matrices $X,Y$ satisfy $\id_{2n}-XX^T\geq 0$, $[X,\sigma_{2n}]=0$, $\kernel(Y)=\kernel(\id_{2n}-XX^T)$ and $2l\geq \rank(\id_{2n}-XX^T)$. 
\label{it:enumsecond}
\end{enumerate}
\end{theorem}
We defer the proof of this theorem to Section~\ref{sec:thmproof}, and first discuss some examples. 
\begin{remark} \label{ref:rankcondv}
Note that if $[X,\sigma_{2n}]=0$, then $\rank(\id_{2n}-XX^T)$ is even (see Lemma \ref{lem:commute}) and therefore also~$\rank(Y)$.
\end{remark}
\begin{example}
Consider the {\em classical noise channel}
given by $X=\id$ and $Y\geq 0$, $Y\neq 0$. 
According to Theorem~\ref{thm:dilation}, this channel is not passively dilatable  because the condition $\kernel(Y)=\kernel(\id-XX^T)$ is  not met. A dilation of this channel with two environment modes is given in~\cite{hol07}.
\end{example}

\begin{example}\label{ex:beamsplitterexample}
Let $U_\lambda$ be the two-mode beamsplitter with transmissivity $\lambda\in [0,1]$, i.e., the Gaussian unitary given by the symplectic matrix
\begin{align}
S_\lambda &=\begin{pmatrix}
\sqrt{\lambda}I_2 & \sqrt{1-\lambda}I_2\\
\sqrt{1-\lambda}I_2& -\sqrt{\lambda}I_2
\end{pmatrix}\label{eq:beamsplitterdef}
\end{align}
with respect to the ordering $(Q_1,P_1,Q_2,P_2)$ of the modes. 
Let $\rho_E$ be a one-mode Gaussian state with covariance matrix~$\gamma_E$. Consider a channel of the form
\begin{align}
\Phi(\rho)&=\tr_E U_\lambda (\rho\otimes\rho_E)U_\lambda^*\ . \label{eq:additivenoisechannelonemode}
\end{align}
We call this an {\em additive }Gaussian channel.

Since $U_\lambda$ is passive, $\Phi$ is clearly passively dilatable. To see that the conditions of the theorem are satisfied, observe that
\begin{align}
X=\sqrt{\lambda}\id_{2}\qquad\textrm{ and }\qquad Y=(1-\lambda)\gamma_E\ .
\end{align}
Assume that $\lambda \in ]0,1[$. Then it is easily verified  (using
the fact that covariance matrices are positive definite) 
that
the conditions of~\eqref{it:enumsecond} are satisfied for any $l\geq 1$. In particular, the theorem implies that there  is a dilation with~$l$ modes for all~$l\geq 1$. This is consistent with expression~\eqref{eq:additivenoisechannelonemode}. The theorem also implies that at least one environment mode is necessary.

On the other hand, assume that $\lambda=1$. Then 
the conditions of~\eqref{it:enumsecond} are satified for any $l\geq 0$, implying the existence of a dilation with no environment modes. Indeed, in this case, the channel is simply the identity channel, with trivial dilation $\Phi(\rho)=\rho$  for all states~$\rho$.

Finally, consider the case where $\lambda=0$. Here the conditions~\eqref{it:enumsecond} apply with $l\geq 1$,  which is also consistent with~\eqref{eq:additivenoisechannelonemode}.

\end{example}

In most cases, the theorem can be stated in a simpler fashion.
\begin{corollary} \label{cor:easycor}
Let $\Phi_{X,Y}$ be an $n$-mode Gaussian channel such that $X,Y$ and $\id_{2n}-XX^T$ have full rank. Then there exists a passive dilation with $n$ modes if and only if $\id_{2n}-XX^T\geq 0$ and $[X,\sigma_{2n}]=0$. \end{corollary}
In fact, we remark that this Corollary can be shown directly by constructing an orthogonal symplectic unitary from $s_1=X$, $s_2=(\id_{2n}-XX^T)^{1/2}$ and using the covariance matrix $\gamma_E=s_2^{-1}Y(s_2^{-1})^{T}$. 

\subsection{General observations about dilations}
We can now make a first step towards proving the theorem:
\begin{lemma} \label{lem:dilationsystem}
Let $\Phi_{X,Y}$ be an $n$-mode Gaussian channel. Using the notation of equation (\ref{eqn:dilation}), such a Gaussian channel can be passively dilated with $l$ environment modes if and only if there exists a tuple $(s_2,\gamma_E)$ 
with $s_2\in \mathbb{R}^{2n\times 2l}$, $\gamma_E\in \mathbb{R}^{2l\times 2l}$ and $\gamma_E\geq i\sigma_{2l}$ such that
\begin{align}
	\begin{split}
		s_2\sigma_{2l}s_2^T&=\sigma_{2n}-X\sigma_{2n}X^T=:\Sigma\\
		s_2s_2^T&=\id_{2n}-XX^T=:\hat{\Sigma} \\
		s_2\gamma_Es_2^T&=Y \label{eqn:dilationsystem}
	\end{split}
\end{align}
Any dilation satisfies $s_1=X$ or $s_1=-X$. 
\end{lemma}
\proof{Given a passive dilation of the channel with a matrix $S\in Sp(2(n+l))\cap O(2(n+l))$, we know that $X\gamma X^T+Y=s_1\gamma s_1^{T}+s_2\gamma_Es_2^T$ for all $\gamma$ by equation~\eqref{eq:actioncovariancematrix}. Therefore it must hold that $s_1\gamma s_1^T=X\gamma X^T$ for all $\gamma \geq i\sigma_{2n}$ and $s_2\gamma_Es_2^T=Y$, which is the third equation of (\ref{eqn:dilationsystem}). 

In particular we have $s_1s_1^T=XX^T$, hence $s_1O=X$ for some orthogonal matrix $O\in O(2n)$ and we have $s_1\gamma s_1^T=s_1O\gamma O^Ts_1^T$ for all covariance matrices $\gamma\geq i\sigma$. This is equivalent to 
\begin{align}
	Q\gamma Q = QO\gamma O^TQ
\end{align} if $Q=s_1^+s_1$ denotes the projection onto $\ker(s_1)^{\perp}$. Since this holds for all covariance matrices $\gamma$ it also holds for all symmetric matrices, in particular the orthogonal projection $Q$. Since $OQO^T$ is a projection itself $QOQO^TQ=Q$ can only hold if $OQO^T=Q$. This implies that in the basis where $Q$ is diagonal, $O$ must be block diagonal and we can write $O=\tilde{O}_1\oplus \tilde{O}_2$ with $\tilde{O}_1$ a matrix onto $\ker(s_1)^{\perp}$ and $\tilde{O}_2$ onto $\ker(s_1)$. 

Since $O$ commutes with $Q$, we also find $Q\gamma Q = O(Q\gamma Q)O^T$ and $O$ commutes with all symmetric matrices $A$ with $\ker(A)\supseteq \ker(s_1)$. This implies that $\tilde{O}_1$ must be a multiple of the identity on $\ker(s_1)^{\perp}$. Since $s_1O=s_1(\tilde{O}_1\oplus 0)$ by construction, this implies that we can find $O^{\prime}\in O(2n)$ such that $s_1O=s_1O^{\prime}$ and $O^{\prime}$ is a multiple of the identity. Since $O^{\prime}$ is orthogonal, $O^{\prime}=\pm \id$. Hence $s_1\gamma s_1^T=X\gamma X^T$ for all $\gamma \geq i\sigma$ if and only if $s_1=\pm X$. 

In addition, we need that $S$ is symplectic and orthogonal, which means that the following conditions must always hold:
\begin{align*}
	s_1\sigma_{2n}s_1^T+s_2\sigma_{2l}s_2^T&=\sigma_{2n} \\
	s_1s_1^T+s_2s_2^T&=\id_{2n} \\
	s_1\sigma_{2n}s_3^T+s_2\sigma_{2l}s_4^T&=0 \\
	s_1s_3^T+s_2s_4^T&=0 \\
	s_3\sigma_{2n}s_3+s_4\sigma_{2l}s_4^T&=\sigma_{2l} \\
	s_3s_3^T+s_4s_4^T&=\id_{2n}
\end{align*}
If we plug in $\pm s_1=X$, the first two conditions are exactly equations (\ref{eqn:orthogextend}) so that it is necessary to satisfy system (\ref{eqn:dilationsystem}) in order to have a passive dilation. 

Conversely, using Lemma~\ref{lem:extension}, having a solution to~\eqref{eqn:dilationsystem}, we can always choose $s_3$ and $s_4$ to extend~$S$ to an orthogonal symplectic matrix.}

This lemma implies that proving Theorem~\ref{thm:dilation} is equivalent to characterizing the solvability of the system of equations (\ref{eqn:dilationsystem}). From the fact that $s_2s_2^T$ is positive semidefinite, it is immediately clear that the system can only be solvable if $\hat{\Sigma}\geq 0$, which is one of the conditions stated in Theorem~\ref{thm:dilation}. To recover the other conditions, we will need the next lemma:

\begin{lemma} \label{lem:commutativity} In the notation of Lemma \ref{lem:dilationsystem}, for any passive dilation of an $n$-mode passively dilatable Gaussian channel $\Phi_{X,Y}$ we have $\Sigma=\sigma_{2n}\hat{\Sigma}$ and both $\Sigma$ and $\hat{\Sigma}$ commute with $\sigma_{2n}$.
\end{lemma}
\proof{By definition, we need $s_2\sigma_{2l}s_2^T=\Sigma$ and $s_2s_2^T=\hat{\Sigma}$. Since $s_2$ is derived from an orthogonal symplectic matrix, it is of the form (see Lemma \ref{lem:unitisom})
\begin{align*}
	s_2=\begin{pmatrix}{} \operatorname{Re}(u_2) & \operatorname{Im}(u_2) \\ -\operatorname{Im}(u_2) & \operatorname{Re}(u_2) \end{pmatrix}.
\end{align*}
Setting 
\begin{align*}
	\mu&:=\operatorname{Re}(u_2)\operatorname{Re}(u_2)^T+\operatorname{Im}(u_2)\operatorname{Im}(u_2)^T \\
	\nu&:=\operatorname{Im}(u_2)\operatorname{Re}(u_2)^T-\operatorname{Re}(u_2)\operatorname{Im}(u_2)^T,
\end{align*}
we obtain:
\begin{align*}
	s_2s_2^T&=\begin{pmatrix}{} \mu & \nu \\ -\nu & \mu \end{pmatrix}\stackrel{!}{=}\hat{\Sigma} \\
	s_2\sigma_{2l}s_2^T&=\begin{pmatrix}{} -\nu & \mu \\ -\mu & -\nu \end{pmatrix}\stackrel{!}{=}\Sigma
\end{align*}
Since $\Sigma$ and $\hat{\Sigma}$ are of the form specified in Lemma \ref{lem:commute}, they commute with~$\sigma_{2n}$.}

\subsection{Proof of Theorem~\ref{thm:dilation}\label{sec:thmproof}}

\subsubsection{Characterization of passively dilatable channels (\eqref{it:enumfirst}$\Rightarrow$\eqref{it:enumsecond})}
We begin by proving the first part of Theorem \ref{thm:dilation}, namely that the stated conditions are necessary:
\begin{lemma} \label{lem:necessary1}
Let $\Phi_{X,Y}$ be an $n$-mode Gaussian channel. The conditions $\id_{2n}-XX^T\geq 0$, $[X,\sigma_{2n}]=0$ and $2l\geq \rank(\id_{2n}-XX^T)$ are necessary for the existence of a passive dilation of the channel with $2l$ environment modes.
\end{lemma}
\proof{By Lemma \ref{lem:dilationsystem}, in order for a dilation to exist, the system of equations (\ref{eqn:dilationsystem}) must be satisfied. In particular, $s_2s_2^T=\id_{2n}-XX^T$. Due to the fact that $s_2s_2^T$ is positive semidefinite, $\id_{2n}-XX^T$ must be positive semidefinite. In addition, if $s_2\in \R^{2n\times 2l}$, then $\rank(s_2s_2^T)\leq 2l$, which implies that $s_2s_2^T=\id_{2n}-XX^T$ can only have a solution if $\rank(\id_{2n}-XX^T)\leq 2l$. Finally, for a passive dilation we have $S\in Sp(2n)\cap O(2n)$ by definition. The $2$-out-of-$3$ property of the unitary group (Lemma \ref{lem:commute}) then implies $[S,\sigma_{2(n+l)}]=0$ and therefore $[s_1,\sigma_{2n}]=0$. Hence $[X,\sigma_{2n}]=0$ is a necessary condition as $X=s_1$.}

\begin{lemma} \label{lem:necessary2}
Let $\Phi_{X,Y}$ be a Gaussian channel. The condition~$\kernel(\id_{2n}-XX^T)=\kernel(Y)$ is necessary for the existence of a passive dilation of the channel.
\end{lemma}
\proof{We suppose that we have found $(s_2,\gamma_E)$ such that $s_2s_2^T=\id_{2n}-XX^T$ and $\gamma_E\geq i\sigma_{2l}$ such that $s_2\gamma_{E} s_2^T=Y$. First note that for every $y\in \kernel(s_2^T)$ we have $s_2\gamma_{E} s_2^Ty=0$, hence $y\in \kernel(Y)$ or $\kernel(s_2^T)\subseteq \kernel(Y)$. Now, on the other hand
\begin{align*}
	\rank(s_2^T)\geq \rank(s_2\gamma_Es_2^T) \geq \rank(s_2^+s_2\gamma_Es_2^Ts_2^{+T})
\end{align*}
with the pseudoinverse $s_2^+$ (see Appendix \ref{app:pseudoinverse} for definition and basic properties), using that the rank of a product of matrices is always smaller than the rank of its factors. Now note that $s_2^+s_2=Q$ is the orthogonal projection onto the range of~$s_2^T$. Since $\gamma_E\geq i\sigma_{2l}$, one can easily see that~$\gamma_E\geq 0$ has full rank, which means that there is~$\varepsilon>0$ such that $\gamma_E\geq \varepsilon \id_{2l}$. Then we have that $Q\gamma_EQ\geq \varepsilon Q^2=\varepsilon Q$, hence 
\begin{align*}
	\rank(s_2^+s_2\gamma_Es_2^Ts_2^{+T})\geq \rank(Q)=\rank(s_2^T)
\end{align*}
But then, $\rank(Y)=\rank(s_2\gamma_Es_2^T)=\rank(s_2^T)$ and therefore $\kernel(Y)=\kernel(s_2^T)$. Finally, since $\kernel(s_2)=\operatorname{im}(s_2^T)^{\perp}$, $\kernel(s_2s_2^T)=\kernel(s_2^T)$ and hence $\kernel(\id_{2n}-XX^T)=\kernel(Y)$ is a necessary condition.}

Lemmas \ref{lem:necessary1} and \ref{lem:necessary2} show that the conditions stated in Theorem~\ref{thm:dilation} are necessary for a passive dilation to exist. This proves the implication~\eqref{it:enumfirst}$\Rightarrow$\eqref{it:enumsecond}.

\subsubsection{Existence of unitary dilations (\eqref{it:enumsecond}$\Rightarrow$\eqref{it:enumfirst})}
We now consider the converse direction, i.e., we assume that $(X,Y)$ satisfy
the conditions stated in~\eqref{it:enumsecond} of Theorem~\ref{thm:dilation} and show that these are sufficient to imply the existence of a passive dilation (as in~\eqref{it:enumfirst}). 
\begin{lemma} \label{lem:existence}
Let $\Phi_{X,Y}$ be an $n$-mode Gaussian channel satisfying $2l\geq\rank(\id-XX^T)$, $\id_{2n}-XX^T\geq 0$, $\kernel(\id_{2n}-XX^T)=\kernel(Y)$ and $[\sigma_{2n}, X]=0$. 
Then there is a passive dilation with $l$ environment modes.
\end{lemma}
\proof{From the spectral theorem, it is known that if $[A,B]=0$ and $A$ is normal, then also $[P_{\lambda(A)},B]=0$ for any spectral projection~$P_{\lambda(A)}$ of $A$ and therefore $[A^{1/2},B]=0$, where $A^{1/2}$ denotes the unique positive square root of $A$. Define $\hat{\Sigma}=\id-XX^T\geq 0$ and $\Sigma=\sigma_{2n}-X\sigma_{2n} X^T$. Using $[\sigma_{2n},X]=0$, we have $\sigma_{2n} \hat{\Sigma}=\Sigma$ and $\sigma_{2n} \hat{\Sigma}=\hat{\Sigma}\sigma_{2n}$, i.e. $\hat{\Sigma}$ commutes with $\sigma_{2n}$. Therefore 
\begin{align}
[\hat{\Sigma}^{1/2},\sigma_{2n}]=0\label{eq:sigmahtcommut}
\end{align}
and thus (see Lemma~\ref{lem:commute}) the matrix $\hat{\Sigma}^{1/2}$  is of the form
\begin{align}
	\hat{\Sigma}^{1/2}=\begin{pmatrix}{} \mu & \nu \\ -\nu & \mu \end{pmatrix}\ \label{eq:sigmatcommuteform}
\end{align}
and
\begin{align}
\Sigma=\hat{\Sigma}^{1/2}\sigma_{2n}\hat{\Sigma}^{1/2}\ .\label{eq:sigmatwondf}
\end{align} 
Furthermore, by definition of the square root (and since $\hat{\Sigma}$ is symmetric), we have 
\begin{align}
(\hat{\Sigma}^{1/2})^T=\hat{\Sigma}^{1/2}\ .\label{eq:symmetrysquareroot}
\end{align}

We divide the proof into three cases:
\begin{enumerate}

\item\label{it:firstcaseln}
Consider  the case where $l=n$. We proceed by constructing a pair $(s_2,\gamma_E)$
satisfying the conditions of Lemma~\ref{lem:dilationsystem}, implying the existence of a passive dilation of~$(X,Y)$. 

Setting $s_2=\hat{\Sigma}^{1/2}$ we have $s_2s_2^T=\hat{\Sigma}$ and $s_2\sigma_{2n}s_2^T=\Sigma$. Thus the first two conditions of~\eqref{eqn:dilationsystem} (Lemma~\ref{lem:dilationsystem}) are satisfied, and it remains to construct a covariance matrix~$\gamma_E$ satisfying $s_2\gamma_Es_2^T=Y$.  Let $s_2^+$ be the Moore-Penrose pseudoinverse of $s_2$. 
We set
\begin{align}
\gamma_E=s_2^{+}Ys_2^{+\,T}+P_{\kernel(s_2)}\ ,\label{eq:gammeexpression}
\end{align}
 where $P_{\kernel(s_2)}$ is the projection onto $\kernel(s_2)$. Then 
\begin{align}
s_2\gamma_E s_2^T&=s_2s_2^+Ys_2^{+\,T}s_2^T=P_{\im(s_2)}YP^T_{\im(s_2)}\label{eq:stwogammeestwo}
\end{align}
where we used the fact that  $s_2 P_{\kernel(s_2)}=0$
in the first identity and the properties of the Moore-Penrose-pseudoinverse (Lemma~\ref{lem:moorepenrose}) in the second step, and where we denoted the projection onto the range~$\im(s_2)$ of $s_2$ by~$P_{\im(s_2)}$.

Since $\im(Y)=\im(\hat{\Sigma})$
by assumption and $\im(\hat{\Sigma})=\im(s_2s_2^T)\subset\im(s_2)$, we have
$P_{\im(s_2)}Y=Y$
and since $Y=Y^T$ is symmetric, it follows that
\begin{align}
P_{\im(s_2)}YP_{\im(s_2)}^T=Y\ .
\end{align}
Inserting this into~\eqref{eq:stwogammeestwo} yields $s_2\gamma_Es_2^T=Y$, as claimed (cf.~\eqref{eqn:dilationsystem}). 
 
 We next verify that $\gamma_E$ is a valid covariance matrix. This is done using equation~\eqref{eqn:comppos}: we have 
\begin{align} 
	s_2^+Ys_2^{+T}+ P_{\kernel(s_2)}
	&\geq (\hat{\Sigma}^{1/2})^{+}(i\sigma_{2n}-iX\sigma_{2n} X^T)(\hat{\Sigma}^{1/2})^{+\,T}+P_{\kernel(s_2)} \nonumber \\
	&= i(\hat{\Sigma}^{1/2})^{+}\Sigma(\hat{\Sigma}^{1/2})^{+\,T}+P_{\kernel(s_2)} \nonumber \\
	&= i(\hat{\Sigma}^{1/2})^{+}\hat{\Sigma}^{1/2}\sigma_{2n}(\hat{\Sigma}^{1/2})^{T}(\hat{\Sigma}^{1/2})^{+\,T}+P_{\kernel(s_2)} \nonumber \\
	&= iP_{\im(s_2^T)}\sigma_{2n} P^T_{\im(s_2^T)}+P_{\kernel(s_2)} 
\end{align}
where we used~\eqref{eq:sigmatwondf}
in the third step and introduced the  projection $P_{\im(\hat{\Sigma}^{1/2})}$
onto the range of the symmetric matrix~$s_2^T=(\hat{\Sigma}^{1/2})^T$ in the fourth step. Since $s_2$ is symmetric we have
\begin{align}
P_{\kernel(s_2)}=\id_{2n}-P_{\kernel(s_2)^\bot}=\id_{2n}-P_{\im(s_2)}=\id_{2n}-P_{\im(s_2^T)}\ .
\end{align}
Using $\id_{2n}\geq i\sigma_{2n}$, we thus obtain
\begin{align}
	s_2^+Ys_2^{+T}+ P_{\kernel(s_2)}&\geq i P_{\im(s_2)}\sigma_{2n} P^T_{\im(s_2)}+
i (\id-P_{\im(s_2)})\sigma_{2n} (\id-P_{\im(s_2)}^T)=i\sigma_{2n} .
\end{align}
Here we used that $P_{\im(s_2)}$ commutes with $\sigma_{2n}$ as a consequence
of~\eqref{eq:sigmahtcommut} and the fact that it is the projection onto the range of $\hat{\Sigma}^{1/2}$.  This concludes the proof that~\eqref{eq:gammeexpression} defines a valid covariance matrix. 
\item\label{it:extendtrivially}
The claim for $l>n$ then follows immediately by using the established claim for $l=n$: since $2n\geq \rank(\id-XX^T)$, there is
a dilation $\Phi(\rho)=\tr_E(U(\rho\otimes \rho_E)U^*)$ involving $n$~environment modes. For an arbitrary $(l-n)$-mode state~$\rho_{\tilde{E}}$, we then have
\begin{align}
\Phi(\rho)&=\tr_{E\tilde{E}}((U\otimes\id_{\tilde{E}})(\rho\otimes (\rho_E\otimes\rho_{\tilde{E}}))(U\otimes \id_{\tilde{E}})^*)\ ,
\end{align}
providing us with a passive dilation using~$l$ modes.

\item Finally, consider the case $l<n$.
Then we have $\rank(\hat{\Sigma})\leq 2l$ by assumption.  
We can assume that $\rank(\hat{\Sigma})=2l$ without loss of generality (cf. Remark~\ref{ref:rankcondv}), since
otherwise we can proceed as in step~\eqref{it:extendtrivially} to increase the number of environment modes.  

We exploit the form~\eqref{eq:sigmatcommuteform} of $\hat{\Sigma}^{1/2}$. 
Because $\hat{\Sigma}^{1/2}$ is symmetric (cf.~\eqref{eq:symmetrysquareroot}),
we have $\mu^T=\mu$ and $\nu^T=-\nu$, hence the complex matrix  $\hat{\Sigma}^{1/2}_{\C}:=\mu+i\nu$ is  Hermitian. We can thus diagonalise $\hat{\Sigma}^{1/2}_{\C}$ with a unitary $u\in U(n)$, which corresponds (see Lemma \ref{lem:unitisom}) to a matrix $o\in Sp(2n)\cap O(2n)$ such that $u\hat{\Sigma}^{1/2}_{\C}u^{\dagger}$ corresponds to $o\hat{\Sigma}^{1/2}o^T$. In particular, 
\begin{align*}
	o\hat{\Sigma}^{1/2}o^T=\diag(d_1,\ldots, d_{l},\underbrace{0,\ldots,0}_{n-l},d_1,\ldots, d_l,\underbrace{0,\ldots,0}_{n-l})
\end{align*}
This implies that $\hat{\Sigma}^{1/2}o^T$ has the form
\begin{align*}
	\hat{\Sigma}^{1/2}o^T=\begin{pmatrix}{} A & 0_{2n\times (n-l)} & B & 
0_{2n\times (n-l)} \end{pmatrix}
\end{align*}
for two matrices $A,B\in \mathbb{R}^{2n\times l}$. 
We now define $s_2$ to be the matrix where we erase the $2(n-l)$ zero columns, i.e. we choose
\begin{align*}
	s_2=\begin{pmatrix}{} A & B \end{pmatrix}\in\mathbb{R}^{2n\times 2l}\ .
\end{align*}
By construction, this implies that $s_2s_2^T=\hat{\Sigma}$ as before, and since $o^T$ commutes with~$\sigma_{2n}$ (Lemma \ref{lem:commute}), we also have $s_2\sigma_{2l}s_2^T=\sigma_{2n}s_2s_2^T$. Again, $\gamma_E$ is defined as in the case $l=n$ by~\eqref{eq:gammeexpression} and we have a solution to the system (\ref{eqn:dilationsystem}) with $\gamma_E\geq i\sigma_{2l}$ by the same argument as in case~\ref{it:firstcaseln}.
 \end{enumerate}} 

\subsection{Minimal dilations}
In the following, we show that under the assumptions of Corollary~\ref{cor:easycor}, any pair of dilations are related by orthogonal symplectic matrices acting on the environment. More generally, let us define a  {\em minimal dilation} as one with the least number of environment modes. We then have the following uniqueness property of minimal dilations.
\begin{theorem}\label{thm:uniqueness}
Let~$\Phi_{X,Y}$ be a passively dilatable $n$-mode Gaussian channel. Then
\begin{enumerate}[(i)]
\item\label{it:dilationminimalfirst}
A dilation is minimal if and only if $l=\frac{1}{2}\rank(Y)$.  There is a minimal dilation given by the construction of Theorem~\ref{thm:dilation}. 
\item\label{it:dilationminimalsecond}
Let
\begin{align}
	S=\begin{pmatrix}{} s_1 & s_2 \\ s_3 & s_4 \end{pmatrix}\label{eq:Smatrixdilation}
\end{align}
be the orthogonal symplectic matrix 
describing the passive Gaussian unitary 
associated with a minimal dilation. Then $\rank(s_2)=2l=\rank Y$. In particular, $s_2\in\mathbb{R}^{2n\times 2l}$ is injective. 
\item\label{it:uniquenessphixy}
Consider two minimal dilations 
\begin{align}
\Phi_{X,Y}(\rho)&=\tr_E U(\rho\otimes\rho_E)U^*=\tr_E \Up(\rho\otimes\rhop_E)\Up^*\ ,
\end{align}
of~$\Phi_{X,Y}$, where $U,\Up$ are passive Gaussian unitaries on $\cH^{\otimes (n+l)}$.  Then there are two passive Gaussian unitaries $\tilde{V},V$ on $\cH^{\otimes l}$ such that
\begin{align}
\Up=(\id_{\cH^{\otimes n}}\otimes \tilde{V})U(\id_{\cH^{\otimes n}}\otimes V)\qquad\textrm{ and }\qquad \rhop_E=V^*\rho_E V\ .
\end{align} 
\end{enumerate}
\end{theorem}
\noindent 
Note that a statement analogous to~\eqref{it:uniquenessphixy} was given in~\cite[Appendix D]{car08} for general (non-passive) dilations.

\begin{remark}\label{rem:mixeddilation}
Let us compare these statements to the results of~\cite{car08,careigioho11}. 
For a channel $\Phi_{X,Y}$, let $l^{\textrm{mixed}}_{\min}(\Phi_{X,Y})$ denote the minimal number of environment modes such that a dilation with a (potentially mixed) state of the environment exists.  By explicit construction, 
it was shown in~\cite{car08} (see also~\cite[Section 2]{careigioho11}) that
$l^{\textrm{mixed}}_{\min}(\Phi_{X,Y})\leq 2n-\rank(\Sigma)/2$, where~$\Sigma$ is defined by~\eqref{eqn:dilationsystem}.  This result was later improved to 
\begin{align}
l^{\textrm{mixed}}_{\min}(\Phi_{X,Y})&\leq \rank(Y)-\rank(\Sigma)/2\ \label{eq:unconstrainedminmixed}
\end{align}
in~\cite{careigioho11}, and this is conjectured to be optimal (a matching lower bound is not known, but see Remark~\ref{rem:puredilations}).  To compare to our results, assume that $\Phi_{X,Y}$ is passively dilatable. Let
$l^{\textrm{mixed}}_{\min,\textrm{passive}}(\Phi_{X,Y})$ denote the minimal number  of environment  modes such that a dilation with a passive unitary exists. By definition, we clearly have 
\begin{align}
l^{\textrm{mixed}}_{\min}(\Phi_{X,Y})\leq l^{\textrm{mixed}}_{\min,\textrm{passive}}(\Phi_{X,Y})\ .
\end{align}
 According to Theorem~\ref{thm:uniqueness}, we have 
\begin{align}
l^{\textrm{mixed}}_{\min,\textrm{passive}}(\Phi_{X,Y})&=\frac{1}{2}\rank Y\ .\label{eq:minimalYhalf}
\end{align} 
But since~$\rank(Y)\geq \rank(\Sigma)$ (see e.g.,~\cite[Eq.~(10)]{careigioho11} -- this follows immediately from the positivity condition~\eqref{eqn:comppos}), this means that
\begin{align}
l^{\textrm{mixed}}_{\min,\textrm{passive}}(\Phi_{X,Y})&=\rank Y-\frac{1}{2}\rank Y\leq \rank(Y)-\rank(\Sigma)/2\ .
\end{align}
Thus our result is consistent with~\eqref{eq:unconstrainedminmixed}. We emphasize that in contrast to the case where passivity is not imposed on the dilating unitary, the exact minimal number  $l^{\textrm{mixed}}_{\min,\textrm{passive}}(\Phi_{X,Y})$ of environment modes is known, i.e., given by expression~\eqref{eq:minimalYhalf}. 
\end{remark}

\begin{remark}\label{rem:puredilations}
The authors of~\cite{car08,careigioho11} also consider dilations where the state~$\rho_E$ is pure. These are referred to as Stinespring dilations.
Correspondingly, they consider the 
minimal number~$l^{\textrm{pure}}_{\min}(\Phi_{X,Y})$ of environment modes for a Stinespring dilation with a pure Gaussian environment state~$\rho_E$ to exist. Imposing Gaussianity here is crucial to get a non-trivial problem, since any mixed state can be purified with only a single additional mode otherwise.
By definition, we clearly have 
$l^{\textrm{mixed}}_{\min}(\Phi_{X,Y})\leq l^{\textrm{pure}}_{\min}(\Phi_{X,Y})$.
Improving an upper bound of~\cite{car08}, and by providing a new lower bound, the identity
\begin{align}
l^{\textrm{pure}}_{\min}(\Phi_{X,Y})= \rank(Y-i\Sigma)\ 
\end{align}
was shown in~\cite{careigioho11}. We have not considered the analogous question for passive dilations, since our focus is on establishing an equivalence with
additive Gaussian channels (see Theorem~\ref{thm:additivechannelcharact}). At least in one direction, the analysis of~\cite[Appendix B]{careigioho11} should be useful: here the minimal number of modes needed to find a Gaussian purification of a generic multimode Gaussian state is computed. 
\end{remark}

\proof{Proof of Theorem \ref{thm:uniqueness}

Statement~\eqref{it:enumsecond} of Theorem~\ref{thm:dilation} implies that there is a dilation with $l=\frac{1}{2}\rank(Y)$ environment, and this number is minimal. This proves statement~\eqref{it:dilationminimalfirst}.

To prove statement~\eqref{it:dilationminimalsecond}, fix a minimal dilation with  orthogonal symplectic matrix~$S$ and covariance matrix~$\gamma_E$. By~\eqref{it:dilationminimalfirst}, the number of environment modes is~$\ell=\frac{1}{2}\rank Y$, i.e.,  $s_2\in\mathbb{R}^{2n\times \rank(Y)}$ and $\gamma_E\in \mathbb{R}^{\rank Y\times\rank Y}$. By the minimality and~\eqref{eq:covariancematrixxydef}, 
we have~$2l=\rank(Y)=\rank(s_2\gamma_Es_2^T)$, but
since $\gamma_E\geq i\sigma_{2l}$, the covariance matrix~$\gamma_E$ is full rank and it follows that
$\rank(s_2)=2l$.  In particular, this implies that~$s_2\in\mathbb{R}^{2n\times 2l}$ is injective.

Finally, we can prove statement~\eqref{it:uniquenessphixy}: Consider two minimal dilations of $\Phi_{X,Y}$ with orthogonal symplectic matrices
\begin{align}
	S=\begin{pmatrix}{} s_1 & s_2 \\ s_3 & s_4 \end{pmatrix}\qquad\textrm{ and }\qquad 
	S'=\begin{pmatrix}{} s'_1 & s'_2 \\ s'_3 & s'_4 \end{pmatrix}
\end{align}
and covariance matrices~$\gamma_E$ and $\gamma_E'$, respectively.  In particular, $s_2,s_2^{\prime}\in \R^{2n\times 2l}$ 
and 
\begin{align}
s_1=s_1'=X\label{eq:equalitysonesoneprime}
\end{align} by~\eqref{eq:covariancematrixxydef}. Using the orthogonality of $S$ and $S'$ (in the form~\eqref{eqn:orthogextend}) 
therefore gives  
\begin{align}
	s_2s_2^T=s_2^{\prime}s_2^{\prime\,T}\ .\label{eq:stwostwotrans}
\end{align}
Since $s_2$ is injective, $s_2^{+}s_2=\id_{2l}$ by the properties of the pseudoinverse. Multiplying~\eqref{eq:stwostwotrans} from the left by $s_2^+$ therefore gives $s_2^T=s_2^+s_2^{\prime}s_2^{\prime\,T}$ and multiplying this from the right with $s_2^{+\,T}$ yields $s_2^Ts_2^{+\,T}=s_2^+s_2^{\prime}s_2^{\prime\,T}s_2^{+\,T}$ which is equivalent to
\begin{align}
	s_2^+s_2^{\prime}(s_2^+s_2^{\prime})^T=\id_{2l}\ .
\end{align}
Hence 
\begin{align}
s_2^+s_2^{\prime}=:o\in O(2l)\label{eq:orthogonalitydef}
\end{align} is orthogonal. Multiplying Eq.~\eqref{eq:orthogonalitydef}  from the left by $s_2$ and using that $s_2s_2^+=P_{\textrm{range}(s_2)}$
is the projection onto the range of $s_2$ 
  we obtain $P_{\textrm{range}(s_2)}s_2^{\prime}=s_2o$, hence
\begin{align}
s_2^{\prime}=s_2o\ \label{eq:stwoorelationship}
\end{align}
because $P_{\textrm{range}(s_2)}s_2^{\prime}=s_2^{\prime}$.
The latter identity follows from the fact that the images of~$s_2$ and~$s_2^{\prime}$ coincide as a consequence of the assumption $s_2s_2^T=s_2^{\prime}s_2^{\prime T}$ and the fact that $s_2^T$ and $s_2^{\prime T}$ are surjective (since $s_2,s_2'$ are injective, as argued above).  

Furthermore, using the symplecticity condition~\eqref{eqn:orthogextend}, we have
\begin{align}
	s_2\sigma_{2l}s_2^T=s_2^{\prime}\sigma_{2l}s_2^{\prime\,T}=s_2o\sigma_{2l}o^Ts_2^T \label{eqn:minimalx}
\end{align}
Since $s_2$ is minimal it is injective and hence $s_2^T$ is surjective. Because of the injectivity of~$s_2$ and the surjectivity of $s_2^T$,  Eq.~\eqref{eqn:minimalx} implies 
\begin{align*}
	\sigma_{2l}=o\sigma_{2l}o^T\ ,
\end{align*}
i.e., $o$ is  orthogonal symplectic, $o\in O(2l)\cap Sp(2l)$. Similarly, $Y=s_2\gamma_Es_2^T=s_2^{\prime}\gamma_E^{\prime}s_2^{\prime\,T}$ by assumption, we have 
\begin{align}
\gamma_E^{\prime}=o^T\gamma_Eo\ .\label{eq:gammaeconn}
\end{align} using once again the injectivity of $s_2$ and $s_2^{\prime}$ (and correspondingly, the surjectivity of $s_2^T$ and $s_2^T$).

Finally, we claim that $S$ and $S'$ only differ by an orthogonal symplectic matrix applied to the environment modes.
Indeed, it follows from~\eqref{eq:equalitysonesoneprime} and~\eqref{eq:stwoorelationship} that
\begin{align}
S\begin{pmatrix}{}
\id_{2n} & 0_{2n\times 2l}\\
0_{2l\times 2n} & o
\end{pmatrix}
&=
\begin{pmatrix}{}
s_1^{\prime} & s_2^{\prime}\\
s_3^{\prime\prime} & s_4^{\prime\prime}
\end{pmatrix}
\end{align}
for some matrices $s_3^{\prime\prime}\in \mathbb{R}^{2l\times 2n}$ and $s_4^{\prime\prime}\in\mathbb{R}^{2l\times 2l}$.  The second part of Lemma~\ref{lem:extension} thus implies that there is an
orthogonal symplectic matrix $o^{\prime}\in Sp(2l)\cap O(2l)$  acting on the $l$~environment modes such that 
\begin{align}
\begin{pmatrix}{}
 \id_{2n} & 0_{2n\times 2l}\\
 0_{2l\times 2n} & o^{\prime}
\end{pmatrix}S\begin{pmatrix}{}
\id_{2n} & 0_{2n\times 2l}\\
0_{2l\times 2n} & o
\end{pmatrix}&=S^{\prime}\ .\label{eq:symplecticconnect}
\end{align}
Combining~\eqref{eq:symplecticconnect} with~\eqref{eq:gammaeconn} yields the claim.}

\subsection{Passive channels}
To conclude this section, we combine Theorem~\ref{thm:dilation} and Theorem~\ref{thm:uniqueness} to characterize passive channels. The latter are defined by having a dilation with a  passive unitary~$U$
and an environment state~$\rho_E$ which is also passive. Here passivity of a state~$\rho_E$ is defined physically by the condition that~$\rho_E$ is the Gibbs state of a passive Hamiltonian $H$ at some  inverse temperature~$\beta$, i.e., $\rho_E=e^{-\beta H}/\tr(e^{-\beta H})$. Mathematically, 
passivity of a state~$\rho_E$ is equivalent to the statement that its covariance matrix~$\gamma_E$ satisfies
\begin{align}
[\gamma_E,\sigma_{2l}]=0\ \label{eq:gammacommutativitycond}
\end{align}
as argued in~\cite{ler13}. In other words, a passive channel is one which  has no ``hidden'' squeezing: both the system-environment interaction and the state of the environment are associated with passive Hamiltonians. We have the following simple characterization of such channels:
\begin{corollary}
Let $\Phi_{X,Y}$ be a passively dilatable Gaussian channel. Then the following are equivalent:
\begin{enumerate}[(i)]
\item
$[Y,\sigma_{2n}]=0$.\label{it:ycond}
\item
$\Phi_{X,Y}$ is passive.\label{it:passivecond}
\end{enumerate}
\end{corollary}
\proof{Suppose $\Phi_{X,Y}$ is passively dilatable. 
We first remark  that any 
orthogonal symplectic matrix~$S$ as in~\eqref{eq:Smatrixdilation}
satisfies
\begin{align}
s_2\sigma_{2l}=\sigma_{2n}s_2\ .\label{eq:commutativitystwo}
\end{align} 
Indeed, this follows immediately  using the block structure of $S$ and $\sigma_{2(n+l)}=\sigma_{2n}\oplus\sigma_{2l}$ by taking the upper right block matrix of the identity $[S,\sigma_{2(n+l)}]=0$.

We prove the two implications:
\eqref{it:ycond}$\Rightarrow$\eqref{it:passivecond}: Assume that $[Y,\sigma_{2n}]=0$. Consider the minimal dilation constructed 
in Theorem~\ref{thm:uniqueness}, with  orthogonal symplectic matrix~$S$ as in~\eqref{eq:Smatrixdilation} and an environment state of~$\ell$ modes with covariance matrix~$\gamma_E$ given by expression~\eqref{eq:gammeexpression}. According to Theorem~\ref{thm:uniqueness}, $s_2$~is injective, hence~$\kernel(s_2)=\{0\}$ and thus $\gamma_E=s_2^+Ys_2^{+T}$.  We will show that~$\gamma_E$ satisfies~\eqref{eq:gammacommutativitycond}, which implies that $\Phi_{X,Y}$ can be passively dilated with a passive environment state~$\rho_E$.

We use~\eqref{eq:commutativitystwo} to  establish the identity
\begin{align}
\sigma_{2l}s_2^+&=s_2^+\sigma_{2n}P_{\textrm{range}(s_2)}\ .\label{eq:sigmastwoplustoshow} 
\end{align}
Indeed, we have 
\begin{align}
s_2^+\sigma_{2n}P_{\textrm{range}(s_2)}-\sigma_{2l}s_2^+&=
s_2^+\sigma_{2n}s_2s_2^+-\sigma_{2l}s_2^+s_2s_2^+
\end{align}
where we used the fact that $s_2s_2^+=P_{\textrm{range}(s_2)}$ 
and $(s_2^+s_2)s_2^+=P_{\textrm{range}(s_2^T)}s_2^+=s_2^+$
by the properties of the pseudoinverse and the fact that $s_2^T$ is surjective (as $s_2$ is injective). That is,
\begin{align}
s_2^+\sigma_{2n}P_{\textrm{range}(s_2)}-\sigma_{2l}s_2^+&=(s_2^+\sigma_{2n}s_2-\sigma_{2l}s_2^+s_2)s_2^+\\
&=(s_2^+s_2\sigma_{2l}-\sigma_{2l}s_2^+s_2)s_2^+\\
&=(P_{\textrm{range}(s_2^T)}\sigma_{2l}-\sigma_{2l}P_{\textrm{range}(s_2^T)})s_2^+=0
\end{align}
where we used~\eqref{eq:commutativitystwo} in the second step and
the fact that $s_2^T$ is surjective (and thus $P_{\textrm{range}(s_2^T)}=\id_{2l}$) in the last step. This establishes~\eqref{eq:sigmastwoplustoshow}.

We will also need the transpose of~\eqref{eq:sigmastwoplustoshow}, which reads
\begin{align}
s_2^{+T}\sigma_{2l}=P_{\kernel(s_2^T)^{\bot}}\sigma_{2n}s_2^{+T}\label{eq:transposekersigm}
\end{align}
because $P^T_{\textrm{range}(s_2)}=P_{\kernel(s_2^T)^\bot}$. 
We can then compute
\begin{alignat*}{3}
\sigma_{2l}\gamma_E&=\sigma_{2l}s_2^+Ys_2^{+T} &&\\
&=s_2^+\sigma_{2n}P_{\textrm{range}(s_2)}Ys_2^{+T}&&\qquad\textrm{by~\eqref{eq:sigmastwoplustoshow}}\\
&=s_2^+\sigma_{2n}Ys_2^{+T}&&\qquad\textrm{because $Y=s_2\gamma_E s_2^T$ }\\
&=s_2^+Y\sigma_{2n}s_2^{+T}&&\qquad\textrm{by the assumption $[Y,\sigma_{2n}]=0$}\\
&=s_2^+YP_{\kernel(s_2^T)^\bot}\sigma_{2n}s_2^{+T}&&\qquad \textrm{since $Y=s_2\gamma_Es_2^T$}\\
&=s_2^+Ys_2^{+T}\sigma_{2l} &&\qquad\textrm{by \eqref{eq:transposekersigm}}\\
&=\gamma_E\sigma_{2l}\ .
\end{alignat*}
Thus $[\gamma_E,\sigma_{2l}]=0$, as claimed.

\eqref{it:passivecond}$\Rightarrow$\eqref{it:ycond}:
 Suppose $\Phi_{X,Y}$ is passive.
Assume  $S$
is an orthogonal symplectic matrix and
$\gamma_E$ a covariance matrix of  a passive state such that
$S$ and $\gamma_E$ define a dilation of the channel $\Phi_{X,Y}$. 
Then $Y=s_2\gamma_Es_2^T$ and thus
\begin{alignat*}{3}
\sigma_{2n}Y&=s_2\sigma_{2l}\gamma_Es_2^T&&\qquad\textrm{ by~\eqref{eq:commutativitystwo}}\\
&=s_2\gamma_E\sigma_{2l}s_2^T&&\qquad\textrm{ because $\rho_E$ is passive, that is,~\eqref{eq:gammacommutativitycond}}\\
&=s_2\gamma_E s_2^T\sigma_{2n}&&\qquad\textrm{ by the transpose of~\eqref{eq:commutativitystwo}}\\
&=Y\sigma_{2n}\ ,
\end{alignat*}
hence $[Y,\sigma_{2n}]=0$ as claimed.}

\section{Passively dilatable channels are additive noise channels}
Consider a (one-mode) channel of  the form 
\begin{align}
\Phi(\rho)= V\left(\tr_E U_\lambda (W\rho W^*\otimes\rho_E) U_\lambda^*\right)V^*\ , 
\end{align}
where~$U_\lambda$
is the beamsplitter of transmissivity~$\lambda$ (see Example~\ref{ex:beamsplitterexample}) and $V,W$ are passive Gaussian (one-mode) unitaries.  That is, $\Phi$ is obtained by applying passive unitaries to the input and output of an additive Gaussian channel. Since $\Phi(\rho)=\tr_E (U(\rho\otimes\rho_E)U^*)$
for  $U=(V\otimes\id_E)U_\lambda (W\otimes \id_E)$, this channel is passively dilatable. Here we show the converse: any passively dilatable is equivalent (up to passive unitaries) to a (multi-mode) additive nois Gaussian channel. The following result is illustrated in Fig.~\ref{fig:additivenoisechannel}.

\begin{figure}
  \centering{
    \begin{tikzpicture}
		\pgfdeclarelayer{bg}    
		\pgfsetlayers{bg,main}

    \node at (-.5,0) (q1) {$A_1$};
		\node at (-.5,-.4) (int1) {\vdots};
    \node at (-.5,-1) (q2) {$A_n$};
    \node at (2,-2) (q3) {$E_1$};
		\node at (2,-2.4) (int2) {\vdots};		
    \node at (2,-3) (q4) {$E_n$};
		\node at (1.2,0.5) (phi) {$\Phi$};
		\draw [decorate,decoration={brace,amplitude=10pt},xshift=-4pt,yshift=0pt] (1.8,-3.3) -- (1.8,-1.7) node [black,midway,xshift=-0.6cm] {\footnotesize $\rho_E$};

		\node[rectangle, draw=black, fill=white, minimum height=1.6cm, minimum width=.8cm] at (3,-.5) (w) {$W$};
		\node[rectangle, draw=black, fill=white, minimum height=1.6cm, minimum width=.8cm] at (7,-.5) (v) {$V$};
		\node[rectangle, draw=black, rounded corners=3pt,minimum height=4.5cm, minimum width=7.5cm,thick] at (4.5,-1.25) (channel) {};

		\begin{pgfonlayer}{bg}
			\fill[black] (4.6,-.95) rectangle (5.4,-1.05);
			\fill[black] (4.6,-1.95) rectangle (5.4,-2.05);

			\draw[-] (0,0) -- (4,0) -- (6,-2) -- (8,-2);
			\draw[-] (0,-1) -- (4,-1) -- (6,-3) -- (8,-3);
			\draw[-] (2.5,-2) -- (4,-2) -- (6,0) -- (9,0);
			\draw[-] (2.5,-3) -- (4,-3) -- (6,-1) -- (9,-1);
		\end{pgfonlayer}				
  \end{tikzpicture}
  }
  \fcaption{This figure shows how a general passively dilatable channel can be understood as an additive noise Gaussian channel composed with passive unitaries (two modes are drawn completely). This defines a normal form of passively dilatable channels.} \label{fig:additivenoisechannel}
\end{figure}

\begin{theorem}\label{thm:additivechannelcharact}
Let $\Phi:\mathcal{B}(A_1\ldots A_n)\rightarrow\mathcal{B}(A_1\cdots A_n)$  be a passively dilatable $n$-mode Gaussian channel.
Then there is an $n$-mode Gaussian state $\rho_E=\rho_{E_1\cdots E_n}$, $n$-mode Gaussian unitaries $V$, $W$  
and transmissivities $\lambda=(\lambda_1,\ldots,\lambda_n)\in [0,1]^n$ such that for the multi-mode beamsplitter $U_\lambda=U^{A_1E_1}_{\lambda_1}\otimes \cdots\otimes U^{A_nE_n}_{\lambda_n}$, we have 
\begin{align}
\Phi(\rho)&= V\left(\tr_E U_\lambda (W\rho W^*\otimes\rho_E) U_\lambda^{*}\right)V^*\qquad\textrm{ for all states } \rho\ .
\end{align}
\end{theorem}
\proof{Assume that $\Phi=\Phi_{X,Y}$ is specified by the pair~$(X,Y)$ of matrices. As in the proof of Theorem~\ref{thm:dilation}, consider $l=n$.
Let $(S,\gamma_E)$ be the dilation constructed in case~\ref{it:firstcaseln}
of the proof of the theorem, i.e.,
$S=\begin{pmatrix}
s_1 & s_2\\
s_3 & s_4
\end{pmatrix}$ satisfies
\begin{align}
s_1=X\qquad\textrm{ and }\qquad s_2=\hat{\Sigma}^{1/2}=(\id-XX^T)^{1/2}\label{eq:sonetwodef}\ 
\end{align} and  the covariance matrix~$\gamma_E$ is given by the expression~\eqref{eq:gammeexpression}. 
Since $[X,\sigma_{2n}]=0$, we can decompose $X$ as  in Lemma~\ref{lem:commute}. Let $D=(G_1+iG_2)(X_1+iX_2)(F_1+iF_2)$ be the singular value decomposition of the complex matrix~$X_1+iX_2$. The matrix~$D$ is nonnegative but not necessarily full rank. 
By definition  and the isomorphism of Lemma~\ref{lem:unitisom}, the unitaries~$G_1+iG_2$ and $F_1+iF_2$ define passive symplectic elements $F,G\in Sp(2n)\cap O(2n)$. 
Define
\begin{align}
\tilde{S}=
\begin{pmatrix}
G & 0\\
0 & \id_{2n}
\end{pmatrix}S\begin{pmatrix}
F & 0\\
0 & G^T
\end{pmatrix}=
\begin{pmatrix}
Gs_1F & Gs_2 G^T\\
s_3F & s_4G^T
\end{pmatrix}=:\begin{pmatrix}
\tilde{s}_1 & \tilde{s}_2\\
\tilde{s}_3 &\tilde{s}_4\ .
\end{pmatrix}\label{eq:tildesvdef}
\end{align}
With~\eqref{eq:sonetwodef} we obtain
\begin{align}
\begin{matrix}
\tilde{s}_1 &=&GXF&=&D\oplus D\\
\tilde{s}_2 &=&G(\id_{2n}-XX^T)^{1/2}G^T&=&(\id_{2n}-D^2\oplus D^2)^{1/2}\ .
\end{matrix}\label{eq:tildesexplicit}
\end{align}
Here we exploited that $XX^T$ is equivalent to $(X_1+iX_2)(X_1+iX_2)^\dagger=(G_1+iG_2)^\dagger D^2 (G_1+iG_2)$ under the isomorphism and hence~$\id_{2n}-XX^T=G^T(\id_{2n}-D^2\oplus D^2)G$. Since $G$ is orthogonal we have $(\id_{2n}-XX^T)^{1/2}=G^T(\id_{2n}-D^2\oplus D^2)^{1/2}G$. 

We conclude from~\eqref{eq:tildesvdef} that
\begin{align}
s_1=G^T\tilde{s}_1F^T\qquad\textrm{ and }\qquad s_2=G^T\tilde{s}_2 G\ ,
\end{align}
i.e., the action of the channel on a covariance matrice~$\gamma$ is given by (cf.~\eqref{eq:actioncovariancematrix})
\begin{align}
X\gamma X^T+Y &=G^T\tilde{s}_1F^T\gamma F\tilde{s}_1^TG+G^T\tilde{s}_2G\gamma_E G^T\tilde{s}_2^T G\ .
\end{align}
Clearly, this means that the channel can be written as the  composition
\begin{align}
\Phi=\Phi_{s_1,s_2\gamma_Es_2^T}=\Phi_{G^T,0}\circ \Phi_{\tilde{s}_1,\tilde{s}_2\tilde{\gamma}_E\tilde{s}_2^T}\circ \Phi_{F^T,0}\ ,
\end{align}
where $\tilde{\gamma}_E=G\gamma_E G^T$ is a valid covariance matrix. It is clear from~\eqref{eq:tildesexplicit} and
the fact that $(\tilde{S},\tilde{\gamma}_E)$ give a dilation 
that $\Phi_{\tilde{s}_1,\tilde{s}_2\tilde{\gamma}_E\tilde{s}_2^T}$ is an additive noise channel, hence the claim follows.}

\subsubsection*{Acknowledgements}
RK is supported by the Technische Universit\"at M\"unchen - Institute for Advanced Study, funded by the German Excellence Initiative and the European Union Seventh Framework Programme under grant agreement no.~291763. He also gratefully acknowledges support by DFG project no.~KO5430/1-1. MI is supported by the Studienstiftung des deutschen Volkes.

\renewenvironment{thebibliography}[1]
        {\frenchspacing
         \small\rm\baselineskip=11pt
         \begin{list}{\arabic{enumi}.}
        {\usecounter{enumi}\setlength{\parsep}{0pt}     
         \setlength{\leftmargin}{17pt}  
                \setlength{\rightmargin}{0pt}
         \setlength{\itemsep}{0pt} \settowidth
          {\labelwidth}{#1.}\sloppy}}{\end{list}}
\nonumsection{References}


\noindent
\begin{thebibliography}{10}

\bibitem{braun05}
Samuel~L. Braunstein.
\newblock Squeezing as an irreducible resource.
\newblock {\em Phys. Rev. A}, 71:055801, May 2005.

\bibitem{car08}
Filippo Caruso, Jens Eisert, Vittorio Giovannetti, and Alexander~S. Holevo.
\newblock Multi-mode bosonic {G}aussian channels.
\newblock {\em New Journal of Physics}, 10(8):083030, 2008.

\bibitem{careigioho11}
Filippo Caruso, Jens Eisert, Vittorio Giovannetti, and Alexander~S. Holevo.
\newblock Optimal unitary dilation for bosonic {G}aussian channels.
\newblock {\em Phys. Rev. A}, 84:022306, Aug 2011.

\bibitem{eis07}
Jens Eisert and Michael~M. Wolf.
\newblock Gaussian quantum channels.
\newblock In N.~J. Cerf, G.~Leuchs, and E.~S. Polzik, editors, {\em Quantum
  Information with continuous variables of atoms and light}. World Scientific,
  USA, 2007.

\bibitem{hol07}
Alexander~S. Holevo.
\newblock One-mode quantum {G}aussian channels: Structure and quantum capacity.
\newblock {\em Problems of Information Transmission}, 43(1):1--11, 2007.

\bibitem{idellercherwolf}
Martin Idel, Daniel Lercher, and Michael~M. Wolf.
\newblock An operational measure for squeezing.
\newblock arXiv:1607.00873.

\bibitem{ler13}
Daniel Lercher, G{\'e}za Giedke, and Michael~M. Wolf.
\newblock Standard super-activation for {G}aussian channels requires squeezing.
\newblock {\em New Journal of Physics}, 15(12):123003, 2013.

\bibitem{mcd98}
Dusa McDuff and Dietmar Salamon.
\newblock {\em Introduction to Symplectic Topology}.
\newblock Oxford Science Publications, 1998.

\bibitem{reckzeilinger}
Michael Reck, Anton Zeilinger, Herbert~J. Bernstein, and Philip Bertani.
\newblock Experimental realization of any discrete unitary operator.
\newblock {\em Phys. Rev. Lett.}, 73:58--61, Jul 1994.

\bibitem{Shannon48}
Claude~E. Shannon.
\newblock A mathematical theory of communication.
\newblock {\em The Bell System Technical Journal}, 27:379--423, 623--656,
  October 1948.

\bibitem{weedbrock2012}
Christian Weedbrook, Stefano Pirandola, Ra\'ul Garc\'{\i}a-Patr\'on, Nicolas~J.
  Cerf, Timothy~C. Ralph, Jeffrey~H. Shapiro, and Seth Lloyd.
\newblock Gaussian quantum information.
\newblock {\em Rev. Mod. Phys.}, 84:621--669, May 2012.

\end{thebibliography}

\appendix

\label{app:pseudoinverse} \noindent
In this appendix, we collect a few well-known facts about the Moore-Penrose pseudoinverse. Let $A\in \R^{k\times m}$ be a not necessarily invertible matrix. Using the singular value decomposition, we can find unitaries $U\in U(k),V\in U(m)$ and a diagonal matrix $D\in \R^{k\times m}$ with $A=UDV$. Define $A^+=V^{\dagger}D^+U^{\dagger}$ with $D^{+}\in \R^{m\times k}$ and $D_{ii}^+=\frac{1}{D_{ii}}$ for all $D_{ii}\neq 0$ and zero otherwise. Then $A^+$ is called the {\em Moore-Penrose pseudoinverse}.
\begin{lemma}\label{lem:moorepenrose}
Let $A\in \R^{k\times m}$ and let $A^+$ be its pseudoinverse. Then:
\begin{enumerate}
	\item $P=AA^+$ is the orthogonal projection onto the range of $A$.
	\item $Q=A^+A$ is the orthogonal projection onto the range of $A^T$.
\end{enumerate}
\end{lemma}

A proof can be found in any introductory book on linear algebra. 

\end{document}